\documentclass[12pt]{article}

\newcommand{\be}{\begin{eqnarray}}
\newcommand{\ee}{\end{eqnarray}}
\newcommand \beq{\begin{eqnarray}}
\newcommand \eeq{\end{eqnarray}}

\def\simge{\mathrel{%
   \rlap{\raise 0.511ex \hbox{$>$}}{\lower 0.511ex \hbox{$\sim$}}}}
\def\simle{\mathrel{
   \rlap{\raise 0.511ex \hbox{$<$}}{\lower 0.511ex \hbox{$\sim$}}}}
\def\bigs{\mathrel{
   \rlap{\raise 0.531ex \hbox{$>$}}{\lower 0.531ex \hbox{$<$}}}}

\begin{document}
\title{ {\bf  An influence functional for ultrasoft QCD} }
\author{F. Guerin$^+$    and E. Iancu$^*$ \\
  $^+$Institut Non Lineaire de Nice,  1361 route des Lucioles,
06560 Valbonne, France
\\  $^*$Service de Physique Th\'eorique, CE-Saclay, 91191 Gif sur
Yvette, France}
\date{}
\maketitle
\begin{abstract}
A real-time path integral for ultrasoft QCD is formulated. It exhibits a
Feynman's influence functional. The statistical properties of the  
theory and  the gauge symmetry are explicit. The  
correspondence is established with the alternative version, where a  
noise term enters a transport equation. \end{abstract}

Ultrasoft QCD is an effective theory for gluons at high temperature  
$T\ \ (g(T)\ll1)$  at distance and time scales larger than  
$(gT)^{-1}$, i.e. the scales $T^{-1}$ and $(gT)^{-1}$ have been  
integrated out \cite{Bod,BI4}. The two dominant features are the  
coloured collective excitations of the gluons, and the important  
energy dissipation towards smaller distance scales. \\
This last feature forbids an imaginary time formalism. In the real time  
formalism, an effective action has been found \cite{FG4} which  
exhibits simple and interesting features. Here, the time has to run  
forwards and backwards along the time axis, $A^+$ is the field  
component that lives on the forward branch, and $A^-$ lives on the  
backward branch. The appropriate basis for writing this action is
\begin{equation}
A^{(1)} = (A^+ + A^-)/2 \ \ , \ \  A^{(2)} = A^+ -A^- \ \ ,
\label{a1}
\end{equation}
known as the Keldysh basis \cite{Chou}. The effective action for  
ultrasoft QCD reads
\begin{equation}
S(A^{(1)},A^{(2)}) = S_0(A^{(1)},A^{(2)}) +\  
S_{\mathrm{IF}}(A^{(1)},A^{(2)}),
\label{a2}
\end{equation}
where $S_0$ is the Yang-Mills action in this  basis:
\begin{eqnarray}
S_0(A^{(1)},A^{(2)}) = - \int_X {1\over4}[F^+_{\mu\nu}F^{+\mu\nu}-  
F^-_{\mu\nu}F^{-\mu\nu}] \nonumber \\
=-{1\over 2}\int_X\ (D^{(1)\mu}
A^{(2)\nu}-D^{(1)\nu} A^{(2)\mu})^a  \  (F^{(1)a}_{\mu\nu} +  g
f^{abc}A^{(2)b}_\mu A^{(2)c}_\nu/4 )
\label{a3}
\end{eqnarray}
where $\int_X=\int_{-\infty}^{+\infty}dt \int d^3x $ , $a$ is a  
color index,
 $D_x^{(1)\,ab}=\delta^{ab}\partial_X+ g f^{abc}A^{(1)c} $
is the covariant  derivative built with the field $A^{(1)}$ in the  
adjoint representation. Furthermore:
\be
S_{\mathrm{IF}}(A^{(1)},A^{(2)})=\  -\int_X
A^{(2)\mu}(X) \ j_\mu(X; A^{(1)})  \nonumber \\
+ \, i\int_{X,Y} {1\over2} \ A^{(2)\mu}(X) \ { K}_{\mu \nu}(X,Y;
A^{(1)}) \ A^{(2)\nu}(Y)  \ \ ,\label{a4}
\ee
where the trace over the colour indices is implicitly understood.
$j^\mu(X;A^{(1)})$ is the response function, i.e. the response of  
the plasma to a small external perturbation at the appropriate  
scale.  In this nonlinear theory
\be
 j^{\mu}(X; A^{(1)})= \Pi^{\mu\nu}_{ret}A^{(1)}_\nu  
+{1\over2}\Gamma^{\mu\nu\rho}_{ret} A^{(1)}_\nu A^{(1)}_\rho +  
\cdots  \ .\label{b0}
\ee
$j_{\mu}$ is
expressed in terms of the (retarded) propagator along jagged paths   
in a background  coloured field $A^{(1)}(X)$ \cite{BI1,BI4}
\be
\lefteqn{ j_{\mu}^{a}(X; A^{(1)}) =  m_D^2\Big[- g_{0\mu}
A_0^{(1)a}(X)
 \nonumber} \\ & &
 +\, i \  \int_{v,v'}  \int d^4Y \ v_{\mu}
  G_{ret}^{ab}( X, Y ; A^{(1)} ; {\mathbf v}, {\mathbf v'})\
\partial_{Y_0} (v'\cdot A^{(1)b}(Y)) \ \Big]
\label{b1}
\ee
where $m_D^2=g^2NT^2/3$, $\int_v=\int {d \Omega_v /{4\pi}}$, and  
$G_{ret}$ is the retarded Green function
 defined as the solution to:
\begin{equation}
i\ (v\cdot D^{(1)}_x + \hat{C}) \ G_{ret}( X, Y; A^{(1)} ;{\mathbf  
v}, {\mathbf v'})  =
 \delta^{(4)}(X-Y) \ \delta_{S_2} ({\mathbf v} - {\mathbf v'})
\label{b3}
\end{equation}
\begin{equation}
\hat{C} \  G_{ret}( X, Y; A^{(1)} ;{\mathbf v}, {\mathbf v'})
=\int_{v"}  \ C({\mathbf{v}, \mathbf{v"}})
G_{ret}( X, Y;A^{(1)} ;{\mathbf v"}, {\mathbf v'})
\label{b4}
\end{equation}
with the condition $ G_{ret} ( X, Y)  =  0$  for $ X_0 < Y_0$ .
$v\cdot D_x^{(1)}$
is the covariant drift
operator  in the space-time direction  $v^\mu=(1,{\mathbf v})$
(${\mathbf v}^2=1$). The  collision operator $\hat{C}$ is local in $X$, 
and changes the  ${\mathbf v}$
direction of the drift. It is  symmetric in ${\mathbf v}$ and
${\mathbf v'}$, and has positive eigenvalues of order $g^2T\ln
1/g$, except for a zero mode $\int_v C({\mathbf v},{\mathbf
v'})=0$. Its explicit form can be found in Refs.
\cite{Bod,BI1,AY2,BI4}.

The dissipation comes from $\hat{C}$ and it gives rise to the  
second term in the action. ${K}_{\mu \nu}$ was found to be  
the complement to $j_\mu$ that is needed to fulfill the KMS  
constraints, which exist, for each $n$, on $n$-point amplitudes\cite{FG4}.
\begin{eqnarray}
\lefteqn{{ K}_{\mu \nu}^{a b}(X,Y; A^{(1)}) \,=\, im_D^2 \ T
\nonumber }\\  & &
\int_{v,v'}\Big[\  v_{\mu}G_{ret}^{ab}(X,Y; A^{(1)};{\mathbf v},
{\mathbf v'}) v_{\nu}' \ + \ v_{\nu}G_{ret}^{ba}(Y,X;
A^{(1)};{\mathbf v}, {\mathbf v'}) v_{\mu}' \ \Big]   \ .
\label{b2} \end{eqnarray}
Note that $i\, G_{ret}(X,Y;A;{\mathbf v}, {\mathbf v'})$   is real,  
 therefore  $j_\mu(X;A^{(1)})$ is real, while $ { K}^{\mu  
\nu}(X,Y;
A^{(1)})$  is real and positive. For example
\be
{K}^{\mu\nu}(p_0, {\bf p};A^{(1)}=0)\,=\,-
\frac{T}{p_0} \,\,2\,{\rm Im}\,\Pi_{ret}^{\mu
\nu}(p_0, {\bf p}) \ .\ee
The occurrence of ${K}^{\mu \nu}(X,Y; A^{(1)})$ in the  
action may be interpreted as the extension to a nonlinear case of  
the fluctuation-dissipation relation . \\

As early as 1963, Feynman\cite{FH} introduced the influence  
functional $\exp \, {iS_{\mathrm{IF}}}$. In a path integral  
approach, this factor accounts for the influence of the environment  
upon the system. Feynman examined the case of  gaussian influence  
functionals, i.e.  $S_{\mathrm{IF}}$ quadratic both in $A^+$ and  
$A^-$, and he showed \cite{FH} that the term linear in $A^{(2)}$  
should be real, while the term quadratic in $A^{(2)}$ should be  
imaginary. The action (\ref{a4}) exhibits these properties for the  
$A^{(2)}$ dependence, while it is fully non-linear in the non-abelian  
gauge field $A^{(1)}$. From Feynman's point of view, the modes of  
the QCD plasma
with momenta $p\ll gT$ made up the ``system'', while the role of
the ``environment'' is played by the modes that have been
integrated out, i.e., by the modes  with momenta $p \geq gT$. 
This interpretation will be fully developed somewhere else
\cite{GI}.   
Recent applications of the influence functional \cite{GM,Hu} did not  
consider gauge theories.  

Ultrasoft QCD is in fact a semi-classical theory \cite{Bod,BI4}. 
Indeed the number  
of gluons  is $N_g(p_0\ll T)\sim T/p_0  \ \gg 1$, i.e.  $N_g\approx  
N_g+1$, and the commutation relations' effect should be negligible.  
   $A^{(1)}$ may be identified with a classical field, while  
$A^{(2)}$ describes the thermal fluctuations around it. The physical role of  
the $A^{(2)}$ field is to balance, via its fluctuation, the  
dissipation associated with the collision term, in such a way that  
the field $A^{(1)}$ remains in thermal equilibrium.

The action (\ref{a3})--(\ref{a4}) is invariant under the
gauge transformations of the $A^{(1)}$ field, with $A^{(2)}$
transforming  covariantly,  as well as $j^\mu$  and
$ {K}^{\mu \nu}$,
\begin{equation}
A^{(1)\mu} \rightarrow h\ A^{(1)\mu} \ h^\dagger - (i / g) \  h \
\partial^\mu \ h^\dagger  \ \ \ , \ \ \ A^{(2)\mu}\rightarrow h \
A^{(2)\mu}\  h^\dagger \ \ .
\label{b12}
\end{equation}
Moreover $j_\mu$
and  ${ K}^{\mu \nu}$ are covariantly conserved:
\be
D^{(1)}_\mu j^\mu &=&0,\nonumber \\
D^{(1)}_{x\:\mu}  \,{ K}^{\mu
\nu}&=& D^{(1)}_{y\:\nu}  \,{ K}^{\mu \nu} \,=\,0\,.
\label{b5}
\ee

The generating functional of the $n$-point correlation functions of  
the fields $A^{(1)},A^{(2)}$ at temperature $T$ is the following
path-integral \cite{GI}
\begin{equation}
Z(J_1,J_2)=\int {\mathcal D}A^{(1)} {\mathcal D}A^{(2)}\
\exp i[S(A^{(1)},A^{(2)})+J_2A^{(1)}+J_1A^{(2)}] \ \ (g.f.)
\label{ZJ}   \end{equation}
where $(g.f.)$ is the gauge-fixing factor. Note that this is written
fully in real time: there is no need for an imaginary-time piece
in the contour since the thermalisation of the fields $A^{(1)}$ 
proceeds exclusively via their scattering off the thermal fluctuations 
of the ``medium'' (the hard modes integrated over, and
simulated here by the fields $A^{(2)}$) \cite{GI}. \\
The measure in the path integral (\ref{ZJ})  
is invariant under the transformation (\ref{b12}). The  
normalization of $Z(J_1,J_2)$  is $Z(J_1,\, J_2=0)= {\mathrm{a \  
constant}}$. Indeed, this condition means that, in addition to  
$S(A^{(1)},A^{(2)}=0)\,=\,0$, all the correlation functions,  
involving the field $A^{(2)}$ only,  vanish, and this property is a   
characteristic feature of the Keldysh basis \cite{Chou}.

Moreover, the fields $A^{(2)}$ are typically weak, so their
mutual interactions can be safely neglected \cite{GI}. Indeed, their 
fluctuations are suppressed  by the large imaginary term in the action, which
implies $\overline{ A^{(2)}}\equiv \sqrt{\langle A^{(2)} \, A^{(2)}\rangle} =
 \sqrt{ K^{-1}}$, with $K^{-1}\propto p_0/T\ll 1$.
It is therefore appropriate to also neglect the term cubic in  
$A^{(2)}$ in the Yang-Mills action (\ref{a3}). Once this is done,
the resulting full action has the simplifying
feature to be quadratic in the ``fluctuation'' field  $A^{(2)}$
(but non-linear to all orders in the ``background field''
$A^{(1)}$). Specifically :%(up to the  gauge-fixing terms):
\be\label{SFULL}
S(A^{(1)},A^{(2)})&=& \int_X
A^{(2)\mu}(X) \ \Big(D^{(1)\nu} F^{(1)}_{\nu \mu} -j_\mu(X;A^{(1)})\Big)
\nonumber\\ &{}&\quad
+ i\int_{X,Y} {1\over2} \ A^{(2)\mu}(X) \ { K}_{\mu \nu}(X,Y;
A^{(1)}) \ A^{(2)\nu}(Y)  \ .
\ee
This  action is manifestly invariant under the gauge transformations  
(\ref{b12}), and it has the additional gauge symmetry  
$A^{(1)}_\mu\rightarrow A^{(1)}_\mu \  ,  \ A^{(2)}_\mu \rightarrow  
A^{(2)}_\mu +D_\mu^{(1)}\Lambda(X)$. One may then fix the gauge and  
perform the path integral on the $A^{(2)}$ field. The choice of a  
background-gauge fixing term for $A^{(2)}$, i.e. $i\,(D^{(1)}_\mu  
A^{(2)\mu})^2 /2\lambda$, has the advantage that it preserves the  
gauge symmetry of $A^{(1)}$, while it introduces $\tilde{  
K}_{\mu\nu} = { K}_{\mu \nu}+  D^{(1)}_\mu D^{(1)}_\nu / \lambda$.  
(An alternative gauge choice is the time-axial gauge $A^{(1)}_0  
=0=A^{(2)}_0 $). As a result of the path integration on $A^{(2)}$,   
a {\it real} gaussian weight appears for the quantity  $D^{(1)\nu}  
F^{(1)}_{\nu \mu} -j_\mu(X;A^{(1)})$, which represents the fluctuations
around the classical equation of motion for the field $A^{(1)}$.
 The generating functional of  
the correlation functions of the field $A^{(1)}$ is now written as a  
path integral on the field configurations, with a real weight, as  
in a statistical mechanics approach. \\

 The comparison with the alternative approach to ultrasoft QCD  
(which is in term of two classical equations) makes use of the  
identity
\begin{eqnarray}
\exp({-A^{(2)} K A^{(2)}/2 })  = [{\mathrm{det}} K]^{-1/2}
\int {\mathcal  D}\eta
 \ {\mathrm{exp}}(-\eta K^{-1} \eta /2 -i \eta A^{(2)})
\label{c0}
\end{eqnarray}
where $A^{(2)} K A^{(2)}=A^{(2)}_i K_{ij} A^{(2)}_j$ ,
 ${\mathcal  D}\eta={\mathcal  D}\eta_i(X)$,
with the  gauge choice  $A^{(2)}_0=A^{(1)}_0=0$ . The action  
$S(A^{(1)}, A^{(2)},\eta)$ is now linear in $A^{(2)}$,  and the  
result of the path integration on $A^{(2)}$
is a functional $\delta$--function which enforces an effective equation
of motion for the field $A^{(1)}$ :
\begin{equation}
D^{(1)\nu} F^{(1)}_{\nu i} -j_i(X;A^{(1)}) - \eta_i(X)=0  \ .
\label{c1}
\end{equation}
The current $\eta_i(X)$ is a random variable with, for {\it fixed}  
$A^{(1)}$,  a real Gaussian distribution
with zero average and 2-point correlation function:
\begin{equation}
 \langle   \eta_i^a(X) \ \eta_j^b(Y)  \rangle  
|_{A^{(1)}{\mathrm{fixed}}} \, =\,   {
K}_{i j}^{ab}(X,Y; A^{(1)})\ .
\label{c2}
\end{equation}
In reality, however,  $A^{(1)}$ is related to $\eta$ via the
non-linear equation of motion (\ref{c1}), so that the true statistics
of the coupled
random fields $A^{(1)}$ and $\eta$ is considerably more involved.
Moreover, in the gauge $A^{(2)}_0=A^{(1)}_0=0$, the Coulomb law   
$D^{(1)\nu} F^{(1)}_{\nu 0}-j_0(X;A^{(1)}) - \eta_0(X)=0$ is a  
consequence of Eqs.~(\ref{c1})--(\ref{b5}), with $\eta_0$ defined by  
$\partial_t\eta_0+D_i\eta_i=0$, and with vanishing initial  
conditions at $t=-\infty$. Eq.~(\ref{c2}) is then generalized to
 \begin{equation}
 \langle   \eta_\mu^a(X) \ \eta_\nu^b(Y)  \rangle  
|_{A^{(1)}{\mathrm{fixed}}} \, =\,   {
K}_{\mu\nu}^{ab}(X,Y; A^{(1)})\ ,
\label{c2b}
\end{equation}
i.e., for  {\it fixed} $A^{(1)}$,  ${K}_{\mu\nu}$ is the  
2-point correlation function of the random source $\eta_\mu$. \\

The original formulation of the
effective theory by  B\"odeker \cite{Bod} involves a
set of coupled equations
\begin{equation}
(D^\nu F_{\nu\mu})_a(X)\, =\,m_D^2\int_v \ v_\mu \ W_a(X,{\mathbf v})
\label{c3}
\end{equation}
\begin{equation}
(v\cdot D+\hat{C})^{ab}W_b(X,{\mathbf v})\, = \,{\mathbf v}\cdot
{\mathbf E}^a + \xi^a
(X,{\mathbf v})  \label{c4}
\end{equation}
where $\xi (X,{\mathbf v}) $
 is a gaussian white noise term, with zero average and  the 2-point  
correlation function
\begin{equation}
\langle\langle \xi^a (X,{\mathbf v}) \ \xi^b (Y,{\mathbf v}')
\rangle\rangle\,=\, {2T\over m_D^2} \ C( {\mathbf v},{\mathbf v}') \   
\delta^{(4)}(X-X') \ \delta^{ab} \ \ .\label{xixi}
\end{equation}
where $\langle\langle \cdots \rangle\rangle$ means average over the  
random white noise.
Eq.~(\ref{c4}) may be solved with the help of the retarded Green
function (\ref{b3}). With the initial condition $W=0$ at
$t=-\infty$, the result is
\begin{equation}
m_D^2\int_v v_\mu  W(X,{\mathbf v})= j_\mu(X;A)+ j^{(\xi)}_\mu(X;A)\ ,
\end{equation}
where $j_\mu(X;A)\equiv j_\mu(X;A^{(1)}=A)$ defined in
Eq.~(\ref{b1}) (with ${\mathbf v}.{\mathbf E} = \partial^0(v.A)-v.D  
\, A^0$), and
\begin{equation}
 j^{(\xi)}_\mu(X;A)=m_D^2 \int d^4Z\int_{v,v'} i\  v_\mu
G_{ret}(X,Z;A;{\mathbf v},{\mathbf v}') \ \xi (Z,{\mathbf v}').
\end{equation}
Thus, Eq.~(\ref{c3}) can be rewritten as:
\be
D^\nu F_{\nu\mu}(X)\, =\,j_\mu(X;A)+ j^{(\xi)}_\mu(X;A)\,,
\label{YMusa}
\ee
where the second piece $j^{(\xi)}_\mu$  of the current
is proportional to the random source
$\xi$, and so is itself a random quantity. One notes that,
for {\it fixed} $A$, $j^{(\xi)}_\mu$ has a Gaussian distribution with
zero average and 2-point correlation function which follows from
Eq.~(\ref{xixi}):
\begin{eqnarray}
\langle\langle \,  j^{(\xi)a}_\mu(X;A)  j^{(\xi)b}_\nu(Y;A)  
\,\rangle\rangle |_{A\ {\mathrm{fixed}}}
=  \, { K}_{\mu\nu}(X,Y;A) \ \ ,
\label{c5}
\end{eqnarray}
 with the use of the identity
\be
\hat{G}^T\  2\hat{C}\ \hat{G}=-i\,\hat{G}^T\,\Big(
({\hat{G}}^T)^{-1}+({\hat{G}})^{-1}\Big) \,
{\hat{G}}=-i({\hat{G}}+{\hat{G}}^T) \,, \label{GCG}
\ee
where ${\hat{G}}^T$ is the transpose of  $\hat{G}$ in $X$ space,
colour space, and ${\mathbf v}$ space , i.e. $i\,\hat{G}^T  
=(-v.D+\hat{C})^{-1}$ in the adjoint representation.

Comparing Eqs.~(\ref{YMusa})--(\ref{c5}) to Eqs.~(\ref{c1})--(\ref{c2b}),  
one concludes that  the random current $j^{(\xi)}_\mu(X;A)$ can be
identified with the variable $\eta_\mu(X)$, and  the field $A(X)$  
with $A^{(1)}(X)$.  \\
An attempt to write down a path integral, involving the field  
$A(X)$ only, was made by Arnold \cite{Ar} in the case of the static  
approximation to the $W$ field equation (\ref{c4}), and in the  
$A_0=0$ gauge. \\

An explicit comparison between the two approaches is possible when  
the Green function $G_{ret}(X,Y;A)$ is expanded in powers of $A$, an  
expansion valid for $P\sim g^2T\ln 1/g$. The solution to the  
classical equations (\ref{c3})(\ref{c4}) is then obtained as an  
expansion in powers of the noise $\xi$, while the expansion of  
$S_{\mathrm{IF}}$  (see Eq.~(\ref{a4})\,) in powers of $A^{(1)}$   
produces effective propagators and 1PI vertices involving  $A^{(1)}$  
and $A^{(2)}$. It has been checked that both approaches give the  
same answer for the $n$-point correlation functions of  
$A^{(1)}\equiv A $ : (i) at the tree-level for $n=2,3$ \cite{FG4} and $n=4$, 
(ii) at the  
one-loop level for $n=2$, as it is now detailed. In the field  
theory approach, the result for $\langle  
A^{(1)}(X)A^{(1)}(Y)\rangle$ is, in momentum space (disregarding the  
indices $\mu,\, \nu$)
\be
G_c(P)=\frac{T}{p_0} \Big(  
\frac{1}{P^2-\Pi_r(P)-\Pi_{r\,\mathrm{new}}(P)} - [P\rightarrow  
-P]\, \Big)
\label{c6}
\ee
where $\Pi_r$ is the ultrasoft self-energy (see Eq.(\ref{b0})\,)  
and $\Pi_{r\,\mathrm{new}}$ is the contribution to the self-energy  
arising from one-loop diagrams with loop-momenta $K\sim g^2T\ln1/g$  
and with effective vertices  and propagators  \cite{FG2}. In the  
classical form $\langle\langle A(X)A(Y)\rangle\rangle$, the terms  
quadratic and quartic in the noise $\xi$ lead to the two  
lowest-order terms in an expansion of $\Pi_{r\,\mathrm{new}}$ of the  
right hand side of Eq.~(\ref{c6}). \\

To conclude, this effective theory provides a full explicit example  
of a Feynman influence functional for a nonlinear case \cite{GI},   
in contrast to  other examples where the emphasis is on the   
``backreaction". Note finally that a similar
influence action can be constructed \cite{GI} for the HTL effective 
theory\cite{BI4}, i.e., the theory for the collective dynamics over the
scale $(gT)^{-1}$.

\end{document}